# Sensing of DNA conformation based on change in FRET efficiency between laser dyes

Dibyendu Dey[1], Jaba Saha[1], Arpan Datta Roy[1], D. Bhattacharjee[1],

Sangram Sinha[2], P. K. Paul[3], Santanu Chakraborty[1], Syed Arshad Hussain[1]*

[1]Thin Film and Nanoscience Laboratory, Department of Physics, Tripura University, Suryamaninagar – 799022, Tripura, India

[2]Department of Botany, Tripura University, Suryamaninagar – 799022, Tripura, India

[3]Department of Physics, Jadavpur University, Kolkata 700 032, India

* Corresponding author

Email: sa_h153@hotmail.com, sah.phy@tripurauniv.in

Phone: +91 381 2375317 (O)

Fax: +91 381 2374802 (O)

ABSTRACT

This communication reports the effect of DNA conformation on fluorescence resonance energy transfer (FRET) efficiency between two laser dyes in layer by layer (LbL) self assembled film. The dyes Acraflavine and Rhodamine B were attached onto the negative phosphate backbones of DNA in LbL film through electrostatic attraction. Then FRET between these dyes was investigated. Increase in pH or temperature causes the denaturation of DNA followed by coil formation of single stranded DNA. As a result the FRET efficiency also changed along with it. These observations demonstrated that by observing the change in FRET efficiency between two laser dyes in presence of DNA it is possible to detect the altered DNA conformation in the changed environment.

## 1. Introduction

One of the most important methods for detecting DNA hybridization is by measuring the fluorescence signals. In this method the dye molecules are intercalated into a DNA double helix [1, 2]. But the inherent limitation of this method is the lack of specificity for many particular duplex [3, 4]. Another most important strategy for the detection of DNA hybridization involves fluorescence resonance energy transfer (FRET). FRET between two molecules is an important physical phenomenon, where transfer of energy occurs from an excited donor flurophore to a suitable acceptor flurophore [5, 6]. This technique is very important for the understanding of some biological systems and has potential applications in optoelectronic and thin film devices [7, 8]. Combining FRET with optical microscopy, it is possible to determine the distance between two molecules within nanometers. The main requirements for the FRET to occur are (i) sufficient overlap between the absorption band of acceptor fluorophore and the fluorescence band of donor fluorophore and (ii) both the donor and the acceptor molecule must be in close proximity of the order of 1–10 nm [5, 6]. The intervening of solvent or other macromolecules has little effect on the FRET efficiency [11-13]. If the distance between the donor–acceptor changes then FRET efficiency also changes and accordingly the FRET process can be used to investigate molecular mechanisms [9, 10]. Due to DNA hybridization or denaturation the donor-acceptor moieties are brought closer together or moved further apart and as a result changes occur in the fluorescence intensity of the FRET pair. There are many reports where the detection and characterization of DNA involves FRET process. K. Fujimoto et al reported the detection of target DNAs by excimer-monomer switching of Pyrene using the FRET process [11]. DNA based nanomachine was reported by H. Liu et al using the FRET phenomenon [12]. Also for encrypting messages on DNA strands, various methods have been accomplished [12, 13].

Double stranded DNA is an interesting anionic polyelectrolyte with unique double helix structure whose base sequence controls the heredity of life [15]. DNA, the eternal molecule shows autocatalytic property i.e. self replication. The process involves the partial separation of two individual strands known as denaturation of DNA. The denaturation of DNA is carried out either by increasing the concentration of hydrogen ions (alkaline denaturation) or by raising the temperature of the medium (melting denaturation). There are many more reports where the melting denaturation of DNA has been investigated [14-16] but a comparatively less attention was given towards the alkaline denaturation [17-19]. But it is very important to have a clear idea about the alkaline denaturation as it may give a hint of what happens in the cell in the changed environment. The stability of a double stranded DNA structure is closely related to the environmental pH [19]. It is important to understand the behavior of DNA from the point of view of its function in the cell. Packaging of DNA in the cell relates to gene expression. Of late a lot of efforts have been given to understand the physicochemical behavior of DNA in different conditions. Still a lot of research efforts have to be done in order to understand its behavior from the point of view of its functions in the cell as well as for its technological applications.

The present communication focuses on the changes in FRET efficiencies between two dyes Acriflavine (Acf) and Rhodamine B (RhB) associated in LbL films in presence of double stranded DNA as well as denatured DNA. Acf and RhB, in principle, are suitable for energy transfer. The fluorescence spectrum of Acf overlaps sufficiently with absorption spectrum of RhB. In the present case the denaturation of the DNA has been studied in a wide range of changes in pH and temperature. It is observed that with the increase in both temperature and pH the DNA is first denatured and finally converted into coil shape. As a result the FRET efficiency between Acf and RhB initially decreases due to the separation of strands of DNA (denaturation) and then again the FRET efficiency increases due to the coil formation of the denatured DNA strands.

## 2. Materials and methods

2.1. Materials

Both the dyes Acf and RhB were purchased from Sigma Chemical Co., USA and used as received. Ultrapure Milli-Q water (resistivity 18.2 MΩ-cm) was used as solvent. The dyes used in these studies are positively charged. The DNA used is Salmon sperm DNA, purchased from SRL India and was used as received. The purity of DNA was checked by UV-Vis absorption and fluorescence spectroscopy before use. Poly acrylic acid (PAA) and poly allylamine hydrochloride (PAH) were used as polyanion and polycation during Layer-by-Layer (LbL) film preparation. Both PAA and PAH were purchased from Aldrich Chemical Co., USA and was used without further purification.

2.2. Film preparation

Electrolytic deposition bath of cationic dye RhB and Acf were prepared with $10^{-4}$ M aqueous solution using triple distilled deionized (resistivity 18.2 MΩ-cm) Millipore water. The anionic electrolytic bath of PAA was also prepared with triple distilled deionized Millipore water (0.25 mg/mL). LbL self assembled films were obtained by dipping thoroughly clean fluorescence grade quartz substrate first in the solutions of anionic PAA and then in the mixture of oppositely charged RhB and Acf (50:50 volume ratio) dye. LbL method utilizes the Van der Waals interactions between the quartz slide and PAA as well as electrostatic interaction between PAA and cationic dyes. The details of the method have been discussed elsewhere [20]. The quartz slide was dipped in the aqueous solution of PAA for 30 mins. Then it was taken out and sufficient time was allowed for drying and then rinsing in water bath for 2 minutes so that the surplus anions attached to the surface were washed off. The dried substrate was then immersed in to the cationic dye mixture (RhB + Acf) followed by same rinsing procedure. Deposition of PAA and RhB - Acf layers resulted in one bi - layer of self assembled film. The incorporation of DNA into the LbL film was done with the help of aqueous PAH solution (0.25 mg/mL). The quartz slide was first dipped in to the electrolytic aqueous solution of polycation (PAH) for 30 mins followed by same rinsing in water bath and drying procedure and then dipped into the anionic DNA (con.= 0.25 mg/mL) solution which was again followed by rinsing action in water bath. The slide thus prepared was dipped into the cationic electrolytic solution of RhB and Acf. Due to electrostatic interaction, cationic Acf and RhB were adsorbed onto the negatively charged surface of the DNA in LbL films.

    In case of pH variation of DNA, the quartz slide was first dipped into the electrolytic polycation (PAH) aqueous solution for 30 mins followed by same rinsing in water bath and drying procedure and then dipped into the anionic DNA (con.= 0.25 mg/mL) solution of different pH (5-13.5) values, which is again followed by rinsing action in water bath. NaOH and HCl were used to increase and decrease the pH of DNA solution.

2.3. UV–Vis absorption and fluorescence spectra measurement

UV–Vis absorption and fluorescence spectra were recoded by a Perkin Elmer Lambda-25 Spectrophotometer and Perkin Elmer LS-55 Fluorescence Spectrophotometer respectively. For absorption measurement the LbL films were kept perpendicular to the incident light. The fluorescence light was collected from the sample surface at an angle of $45^0$ (front face geometry) and the excitation wavelength was 420 nm.

2.4. AFM measurement

Atomic force microscopy (AFM) image of one bi-layer LbL film was taken in air with commercial AFM system (Bruker Innova). The AFM image presented here was obtained in intermittent contact (tapping) mode. Typical scan area was $1\times1$ μm$^2$. The Si wafer substrate was used for the AFM measurement.

**3. Results and discussions**

3.1. FRET between Acf and RhB in presence and absence of DNA

In principle Acf and RhB are suitable for FRET [21]. The fluorescence spectrum of Acf sufficiently overlaps with the absorption spectrum of RhB. Both the dyes are highly fluorescent. There are few reports on the investigation of FRET between these two dyes [21, 22]. In our laboratory we have also studied FRET phenomenon using these two dyes [22]. The absorption and fluorescence spectra of Acf and RhB in LbL films (figure S1 of the supporting information) suggest that both the dyes remain as monomers in the LbL films [21, 23]. The corresponding absorption and fluorescence maxima are shown in table 1. Figure 1 shows the fluorescence spectra of Acf-RhB mixed LbL films in presence and absence of DNA. The fluorescence spectra of pure Acf and RhB are also shown in figure 1. All the spectra were recorded with excitation wavelength at 420 nm. The excitation wavelength was selected in order to excite the Acf molecules directly and to avoid any direct excitation of the RhB molecules. With this excitation wavelength, Acf shows strong fluorescence (curve 1 of figure 1), since Acf absorbs light in this excitation range. On the other hand the RhB fluorescence (curve 2 of figure 1) is almost negligible. It is interesting to note that, for Acf - RhB mixed LbL films (curve 3 of figure 1) the RhB fluorescence intensity increases and the Acf fluorescence intensity decreases with respect to their pure counter parts. This is possible if some energy is transferred from Acf to RhB and this transferred energy excites more RhB molecules followed by light emission from RhB. Thus an increase in RhB fluorescence intensity and decrease in Acf fluorescence intensity was found. This possibility has been confirmed by measuring the excitation spectra, where the monitoring emission maxima 525 nm (Acf) and 580 nm (RhB) in case of Acf-RhB mixed LbL films. It has been observed that both the excitation spectra are almost similar (figure S2 of supporting information) and possess characteristic absorption bands of Acf monomers. Thus FRET between Acf and RhB has been confirmed. Further increase in RhB fluorescence and decrease in Acf fluorescence in Acf-RhB fluorescence spectra (curve 4 of figure 1) have been observed for the Acf-RhB mixed LbL films prepared in presence of DNA. This is an indication that the presence of DNA enhances the FRET. It is interesting to mention in this context that in one of our previous work we have investigated the influence of DNA in the FRET between Acf and RhB in solution phase [24]. Based on that investigation we have demonstrated a DNA sensor [24].

The FRET efficiencies were calculated from fig. 1 using the equation given below [25].

$$E = 1 - \frac{F_{DA}}{F_D}$$

Where $F_{DA}$ is the relative fluorescence intensity of the donor in the presence of acceptor and $F_D$ is the fluorescence intensity of the donor in the absence of the acceptor.

It is interesting to note that the FRET efficiency of the dye pair increases from 28.42% to 44.62% in presence of DNA. It is relevant to mention in this context that FRET process is distance dependent and if the inter molecular distance between donor and acceptor pair decreases, then the transfer of energy from donor to acceptor becomes more efficient [5, 6]. The DNA is negatively charged and so the water soluble cationic dyes Acf and RhB are attached electrostatically with the phosphate back bone of DNA resulting a close proximity and favorable condition for efficient energy transfer. This has been shown schematically in the later part of the manuscript.

3.2. Effect of pH on DNA and FRET

In order to investigate the effect of DNA denaturation on the FRET between Acf and RhB, we have prepared DNA LbL films at different pH and Acf-RhB mixed LbL films in presence of DNA at different pH. Figure 2(a) shows the variation of absorbance intensity of the 260 nm band of DNA-PAH LbL films with increasing pH range from 5-13.5. From figure 2(a) it is observed that the absorbance intensity of 260 nm band of DNA-PAH LbL films increases abruptly with increase in pH and almost at pH 11 it becomes maximum. At higher pH the denaturation of DNA occurred resulting in a increase in absorbance intensity [15, 26]. There are several reports on the denaturation and renaturation of DNA with pH variation based on the hypochromatic shift of DNA absorbance [15, 26]. It is interesting to mention in this context that due to the increase in pH or temperature, the H-bond in the double stranded DNA is disrupted and the strands are no longer held together, that is the double helix structure is denatured and the strands separate from each other and finally form individual random coils. The course of this dissociation can be followed spectrophotometrically, as the relative absorbance of the DNA-PAH-LbL films at 260 nm increases with DNA denaturation [15]. UV absorption of the bases in the DNA is due to the π-electron transition. In double stranded DNA the aromatic bases are stacked together and they interact via their p-electron clouds. This results a decrease in π-electron transition probability. As a result the absorption intensity also decreases. Where as in case of single stranded DNA the unstacking of aromatic bases of DNA decrease the interaction between the p-electrons resulting in an increase in π-electron transition probability. As a result the UV absorption intensity increases. At pH values greater than 10, extensive deprotonation of the bases occurs, destroying their hydrogen bonding potential and denaturing the DNA duplex [27]. Above pH 12 the absorbance intensity decreases abruptly indicating the single stranded to coil transition due to the ionization of the guanine moieties of the DNA [27]. The formation of water by the introduction of hydroxyl ions present in the solution with a proton of the H-bridges between the base pairs released some amount of energy during denaturation. This energy is mainly responsible for the coiling of single stranded DNA [28]. On the other hand the length of transition dipole moment vector increases due to the breaking of H-bridges with increasing pH, which results in an increase in absorption intensity. But the formation of coil of the single stranded DNA results a remarkable decrease in the length of the transition dipole moment vector. This causes an abrupt decrease in absorption intensity.

Figure 2(b) shows the variation of FRET efficiencies between Acf and RhB in the Acf-RhB LbL films in presence of DNA at different pH range from 5-13.5. The corresponding FRET efficiencies have also been calculated and shown in table 2. It is observed that the transfer of energy from Acf to RhB in presence of DNA almost remains constant up to pH 9. This implies that the strands are not separated from each other (distance between Acf and RhB are unchanged) up to this pH. But from pH 10-12 the FRET efficiencies decrease remarkably indicating the separation between the strands of the DNA resulting in an increase in the separation between the dye molecules attached electrostatically with the DNA strands. But increasing the pH after 12 results in some amount of increase in FRET efficiency between the Acf-RhB pair. This is because after pH 12 the single stranded DNAs start converting in to coil and as a result the separation between the dye molecules decreases providing a slight increase in FRET efficiency. This situation has been explained schematically in the later part of the manuscript.

3.3. Effect of temperature on DNA and FRET

In order to investigate the effect of DNA denaturation on the FRET between Acf and RhB, we have prepared DNA LbL films at different temperature and Acf-RhB mixed LbL films in presence of DNA at different temperature. DNA monolayer was deposited on a PAH layer at different temperature starting from $30^0$ C to $100^0$ C. The variation of absorbance intensity and FRET efficiency in case of temperature change is very similar to that of pH change. The absorbance intensities of the UV-Vis absorption spectra of the corresponding LbL films are shown in figure 3(a). It is observed that the absorbance increases remarkably for the films prepared at higher temperature, up to $80^0$C (figure 3(a)). The interaction between the p-electrons decreases with the rise in temperature due to unstacking of DNA bases. As a result the absorption intensity maximizes. Also the length of transition dipole moment vector increases due to breaking of H-bridges with increasing temperature [15]. This in turn increases the absorption intensity. The increase in temperature after 80$^0$C results in a remarkable decrease in absorption intensity which is obtained due to coil formation of the totally denatured DNAs [15]. The coil formation of DNA results a remarkable decrease in the length of the transition dipole moment vector. This causes an abrupt decrease in the absorption intensity. In one of our earlier work we have demonstrated the immobilization of denatured single stranded DNA in LbL films [15], where it has been observed that the absorption intensity increases for DNA LbL films prepared up to 80$^0$C and beyond this temperature the absorption intensity decreases. Figure 3(b) shows the variation of FRET efficiencies between Acf and RhB in the Acf-RhB mixed LbL films prepared in presence of DNA at different temperature. The calculated efficiencies are given in table 2. Here also the trend and nature of FRET efficiency with temperature are almost similar to the pH change. Up to 60$^0$C temperature the FRET efficiencies remained unchanged and beyond this temperature the efficiencies started decreasing up to 90$^0$C. After this temperature FRET efficiency again increased slightly. This is due to the denaturation of DNA with temperature and consequent formation of coil of the single stranded DNA with temperature. Information about denaturation and coil formation of the DNA as extracted from figure 2 and 3 are listed in table 3.

3.4. Atomic Force Microscopy study of DNA

In order to check the difference in the surface morphology of the DNA – PAH LbL films before and after melting we have measured the AFM image of the DNA films prepared at different pH and temperature. Figure 4 shows the representative AFM images of DNA films prepared at ambient condition (normal pH and room temperature), at higher pH 13 and higher temperature 95$^0$C. Values of RMS Roughness, roughness average and average heights as extracted from the AFM images are listed in table 4. Height profile analyses along with the line marked on the images are also shown.

From the AFM images it has been observed that almost all the substrate surface was completely covered by the DNA molecules in the backbone of PAH in LbL films. Interesting thing is that AFM images of the LbL films prepared at different temperature and pH values before denaturation (temperature < $80^0$C and pH < 11) show almost similar topology. DNA molecules lying flat on the substrate surface are clearly visible. From the height profile analysis it has been observed that height of both the films varies from 0 to 3.5 nm. Again the AFM images of the DNA films prepared after denaturation at pH 13 and temperature $95^0$C also reveal identical topology. Here DNA molecules do not remain flat on the films, rather, they posses coiled structure. Height profile analysis show that thickness of the films varies within 0–16 nm.

Therefore, the films prepared before and after denaturation show completely different morphology. Considering the diameter of the DNA molecule ~ 2 – 2.5 nm, it seems that the DNA molecules lie flat on the films before denaturation. Again after denaturation the DNA molecules forms coil and possess globular structure. Therefore, AFM investigation gives compelling visual evidence of change in DNA conformation in LbL films before and after denaturation.

3.5. Schematic diagram

The double stranded DNA is composed of two long polymer strands and the strands are attached by hydrogen bonds called nucleotides (or bases). There are four types of bases in DNA, and they are Adenine (A), Thymine (T), Guanine (G) and Cytosine (C), as shown in figure 5(a) and 5(b). The bases lie horizontally between the two spiraling polymer strands. The distance between two consecutive base pairs is 0.34 nm [29-31]. The negatively charged phosphate deoxyribose backbones on either side of the base pair can be labeled with different functional groups or dye molecules [29-31]. A schematic representation showing the attachment of RhB (A) and Acf (D) on the negatively charged phosphate backbone of double stranded DNA is given in figure 5(c). Both the dyes Acf and RhB used in the present study are cationic. In presence of DNA they are attached with the DNA strands through the electrostatic attraction with the negatively charged phosphate backbone of DNA (figure 5c). Before denaturation the FRET efficiency between the dyes adsorbed on to DNA remains almost constant. This is because the distance between them is almost constant. Again after denaturation (at high pH and temperature) the FRET efficiency decreases. After denaturation the DNA strands are separated and remain as single stranded forms resulting an increase in separation between the phosphate backbones and hence the dyes. This has been shown schematically in figure 5(d). However above the pH 12 and temperature $90^0$ C the FRET efficiency again increases slightly. At this stage the single stranded DNA forms coiled structure resulting slight decrease in separation between dyes in compared to the single stranded form. This situation has been shown in figure 5(e).

## 4. Conclusion

Based on the experimental observation it is concluded that the presence of DNA increases the FRET efficiency between two laser dyes Acf and RhB in LbL film. Increase in pH or temperature of DNA results in denaturation followed by coil formation of the DNA. The denaturation of DNA provides an increase in the intermolecular separation between the Acf and RhB molecules resulting a decrease in FRET efficiency. On the other hand the coil formation of the DNA strands provides a decrease in the intermolecular separation between Acf and RhB molecules resulting an increase in FRET efficiency. The present investigation revealed that by observing the change in FRET efficiency between two laser dyes attached to DNA strands, it would be possible to detect the altered DNA conformation in the changed environment.

**Acknowledgements**

The author SAH acknowledge the financial support to carry out this research work through DST Fast-Track project Ref. No. SE/FTP/PS-54/2007, CSIR project Ref. 03(1146)/09/EMR-II.

**References**

[1] V.V. Demidov, Tough nuts to crack: encouraging progress in peptide nucleic acid hybridization to structured DNA/RNA targets, Trends Biotechnol. 21 (2003) 148-51.

[2] F.F. Bier, F. Kleinjung, J. Fresenius, Feature-size limitations of microarray technology--a critical review, Anal. Chem. 371 (2001) 151-156.

[3] G.J. Nuovo, In situ localization of PCR amplified DNA and cDNA, Methods Mol. Biol. 123 (2000) 217-238.

[4] M.A. Lee, A.L. Siddle, R.H. Page, ResonSense: simple linear fluorescent probes for quantitative homogeneous rapid polymerase chain reaction, Anal Chim Acta 457 (2002) 61-70.

[5] T.H. Förster, Experimentelle und theoretische Untersuchung des Zwischenmolekularen übergangs von Elektrinenanregungsenergie, Naturforsch 4A (1949) 321-327.

[6] T.H. Förster, Transfer mechanisms of electronic excitation, Diss. Faraday Soc. 27 (1959) 7–71.

[7] G. Haran, Topical Review: Single-molecule fluorescence spectroscopy of biomolecular folding, J. Phys. Condens. Matter 15 (2003) R1291–R1317.

[8] M.S. Csele, P. Engs, Fundamentals of Light and Lasers, Wiley, New York, 2004.

[9] C.S. Yun, A. Javier, T. Jennings, Fisher, S. Hira, S. Peterson, B. Hopkins, N.O. Reich, G.F. Strouse, Nanometal Surface Energy Transfer in Optical Rulers, Breaking the FRET Barrier, ReV. B 26 (1982) 5409-5413.

[10] B Zagrovic, C.D. Snow, M.R. Shirts, V.S. Pande, Simulation of folding of a small alpha-helical protein in atomistic detail using worldwide-distributed computing, J. Mol. Biol. 323(2002) 927-937

[11] K. Fujimoto, H. Shimizu, M. Inouye, Unambiguous Detection of Target DNAs by Excimer-Monomer Switching of Molecular Beacons, J. Org. Chem. 69 (2004) 3271-3275.

[12] H. Liu, Y. Zhou, Y. Yang, W. Wang, L. Qu, C. Chen, L. Liu, D. Zhang, D. Zhu, Photo-pH Dually Modulated Fluorescence Switch Based on DNA Spatial Nanodevice, J. Phys. Chem. B 112 (2008) 6893-6896.

[13] K. Tanaka, A. Okamotoa, I. Saito, Public key system using DNA as a one-way function for key distribution, BioSystems 81 (2005) 25-29.

[14] J. Marmur, P. Doty, Heterogeneity in deoxyribonucleic acids. II. Dependence on composition of the configurational stability of deoxyribonucleic acids, Nature 183 (1959) 1427-1429.

[15] S.A. Hussain, P.K. Paul, D. Dey, D. Bhattacharjee, S. Sinha, Immobilization of single strand DNA on solid substrate, Chemical Physics letter 450 (2007) 49-54

[16] J. Eigner, P. Doty, The native, denatured and renatured states of deoxyribonucleic acid, J. Mol. Biol. 549 (1965) 549-580.

[17] D.M. Crothers, N.R. Kallenbach, B.H. Zimm, The melting transition of low-molecular-weight DNA, J. Mol. Biol. 802 (1965) 802-820.

[18] F.W. Studier, Sedimentation studies of the size and shape of DNA, J. Mol. Biol. 11 (1965) 373-390.

[19] P.F. Davison, The rate of strand separation in alkali-treated DNA, J. Mol. Biol. 22 (1966) 97-108.

[20] D. Dey, S.A. Hussain, R.K. Nath, D. Bhattacharjee, Preparation and characterization of an anionic dye-polycation molecular films by electrostatic Layer-by-Layer adsorption process, Spectrochimica Acta A 70 (2008) 307-312.

[21] P.D. Sahare, V.K. Sharma, D. Mohan, A.A. Rupasov, Energy transfer studies in binary dye solution mixtures Acriflavine+Rhodamine 6G and Acriflavine+Rhodamine B, Spectrochim. Acta Part A 69 (2008) 1257-1264.

[22] D. Dey, D. Bhattacharjee, S. Chakraborty, S.A. Hussain, Effect of nanoclay laponite and pH on the energy transfer between fluorescent dyes, J. Photochem. Photobiol. A-Chem 252 (2013) 174–182

[23] S.A. Hussain, S. Banik, S. Chakraborty, D. Bhattacharjee, Adsorption kinetics of a fluorescent dye in a long chain fatty acid matrix, Spectrochim. Acta Part A 79 (2011) 1642–1647.

[24] D. Bhattacharjee, D. Dey, S. Chakraborty, S.A. Hussain, S. Sinha, Development of a DNA sensor using a molecular logic gate, Journal of Biological Physics 39 (2013) 387-394.

[25] D. Seth, D. Chakrabarty, A. Chakraborty, N.S. Sarkar, Study of energy transfer from 7-amino coumarin donors to rhodamine 6G acceptor in non-aqueous reverse micelles, Chem. Phys. Lett. 401 (2005) 546-552.

[26] Y.C. Wang, C.B. Lin, J.J. Su, Y.M. Ru, Q. Wu, Z.B. Chen, B.W. Mao, Z. Wu Tian, Electrochemically-Driven Large Amplitude pH Cycling for Acid–Base Driven DNA Denaturation and Renaturation, Anal. Chem. 83 (2011) 4930–4935.

[27] C. Zimmer, Rate of production of single-strand breaks in DNA by JT-irradiation in situ, Biochim. Biophys. Acta. 161 (1968) 584- 586.

[28] M. Ageno, E. Dore, C. Frontali, The alkaline denaturation of DNA, Biophysical Journal 9 (1969) 1281-1311

[29] J. Malicka, I. Gryczynski, R.L. Joseph, DNA hybridization assays using metal-enhanced fluorescence, Biochemical and Biophysical Research Communications 306 (2003) 213-218.

[30] N. Mathur, A. Aneja, P.K. Bhatnagar, P.C. Mathur., Triple-FRET Technique for Energy Transfer Between Conjugated Polymer and TAMRA Dye with Possible Applications in Medical Diagnostics, Journal of Sensors 34 (2008) 487-493.

[31] J.D. Watson, F.H.C. Crick, A structure for Deoxyribose Nucleic Acid, Nature 171 (1953) 737-738.

**Table caption**

**Table 1** The band positions of absorption and fluorescence maxima of Acf and RhB in LbL films.

**Table 2** Variation of FRET efficiencies between Acf and RhB in LbL film in presence of DNA at different pH and temperature

**Table 3** Different physical parameters of Salmon sperm double stranded DNA

**Table 4** Values of different parameters as extracted from the AFM images

**Figure caption**

**Fig.1.** Fluorescence spectra of pure Acf (1), pure RhB (2), Acf+RhB (3) and Acf+RhB with DNA (4) in LbL film.

**Fig.2. (a)** Variation of absorbance intensity of the 260 nm peak of the DNA-PAH LbL film prepared at different pH. **(b)** Variation of FRET efficiency of Acf+RhB LbL film in presence of DNA prepared at different pH.

**Fig.3. (a)** Variation of absorbance intensity of the 260 nm peak of the DNA-PAH LbL film prepared at different temperature. **(b)** Variation of FRET efficiency of Acf+RhB LbL film in presence of DNA prepared at different temperature.

**Fig.5.** Schematic diagram shows **(a)** the four bases of DNA and **(b)** double helix structure of DNA. Attachment of FRET pair (Acf+RhB) with **(c)** DNA double helix structure, **(d)** DNA denatured form and **(e)** DNA coil form.

**Fig.4.** AFM images of the DNA LbL films prepared at **(a)** ambient condition (normal pH and room temperature), **(b)** pH = 13 and **(c)** temperature = $95^0$C.

**Tables**

**Table 1**

| Absorbance | | Fluorescence | |
|---|---|---|---|
| Acf | RhB | Acf | RhB |
| 450 nm (monomer) | 568 nm (monomer) | 525 nm (monomer) | 580 nm (monomer) |
| | 530 nm (0-1 vibronic) | | |

**Table 2**

| pH | FRET efficiency (%) | Temperature ($^0C$) | FRET efficiency (%) |
|---|---|---|---|
| 5 | 44.71 | 30 | 45.60 |
| 6 | 44.62 | 40 | 45.51 |
| 7 | 44.65 | 50 | 45.32 |
| 8 | 44.51 | 60 | 45.12 |
| 9 | 44.35 | 70 | 44.34 |
| 10 | 43.58 | 80 | 37.28 |
| 11 | 36.86 | 90 | 35.44 |
| 12 | 32.45 | 95 | 36.11 |
| 13 | 32.89 | 100 | 38.50 |
| 13.5 | 34.64 | | |

**Table 3**

| DNA | Density (gm/cc) | G-C content (mol %) | Denaturation | Coil formation | Hypochrometic shift |
|---|---|---|---|---|---|
| Salmon sperm | 1.703 | 42 | pH=11 | pH=13 | pH=43% |
| | | | Temp=80$^0$C | Temp=95$^0$C | Temp=47% |

**Table 4**

| | Ambient condition | pH=13 | Temp = 95$^0$C |
|---|---|---|---|
| RMS Roughness (nm) | 0.5993 | 4.1591 | 4.2657 |
| Roughness average (nm) | 0.4839 | 3.3996 | 3.4933 |
| Average height (nm) | 2.9236 | 12.0051 | 13.1076 |

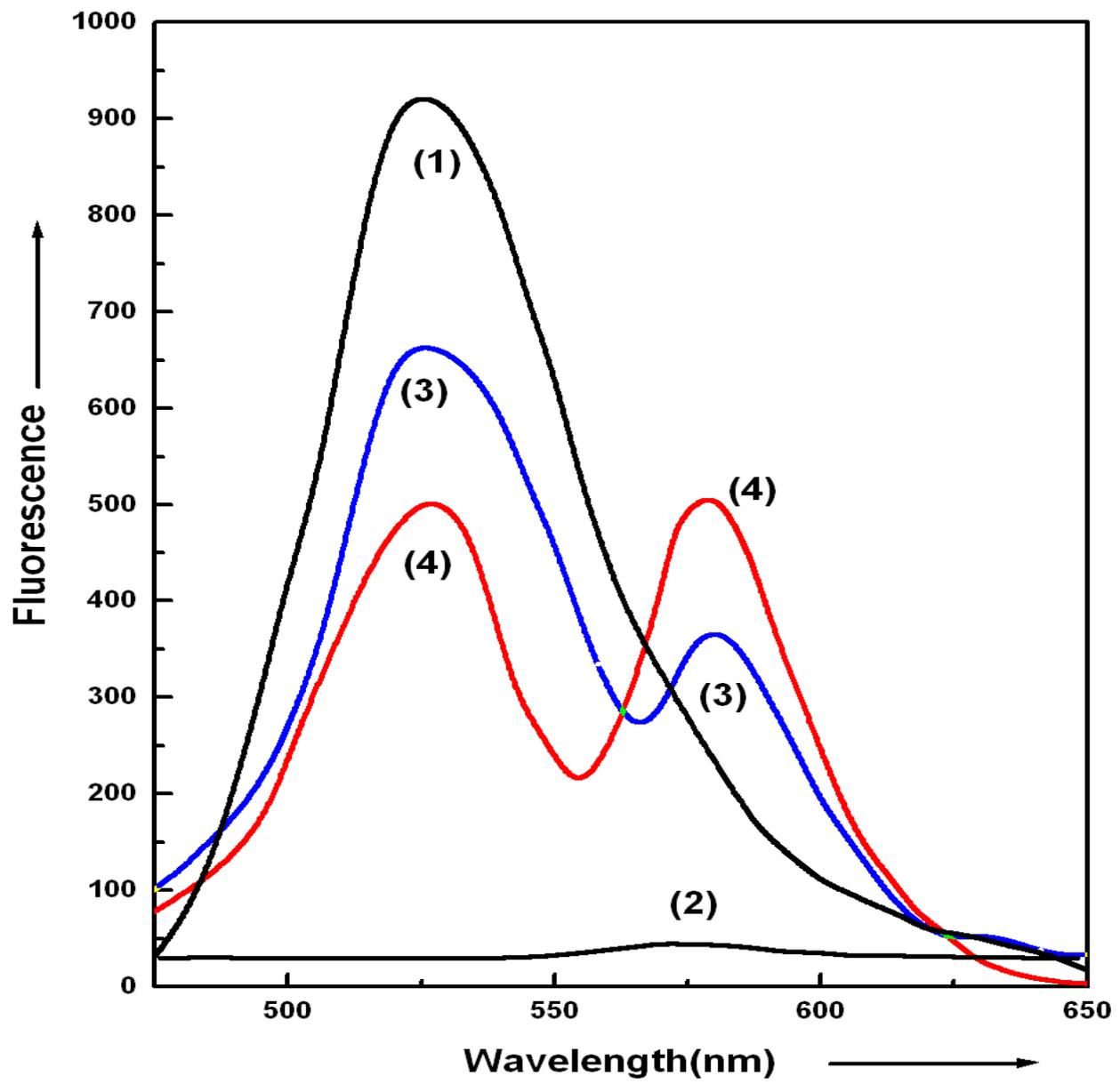

**Fig.1.**

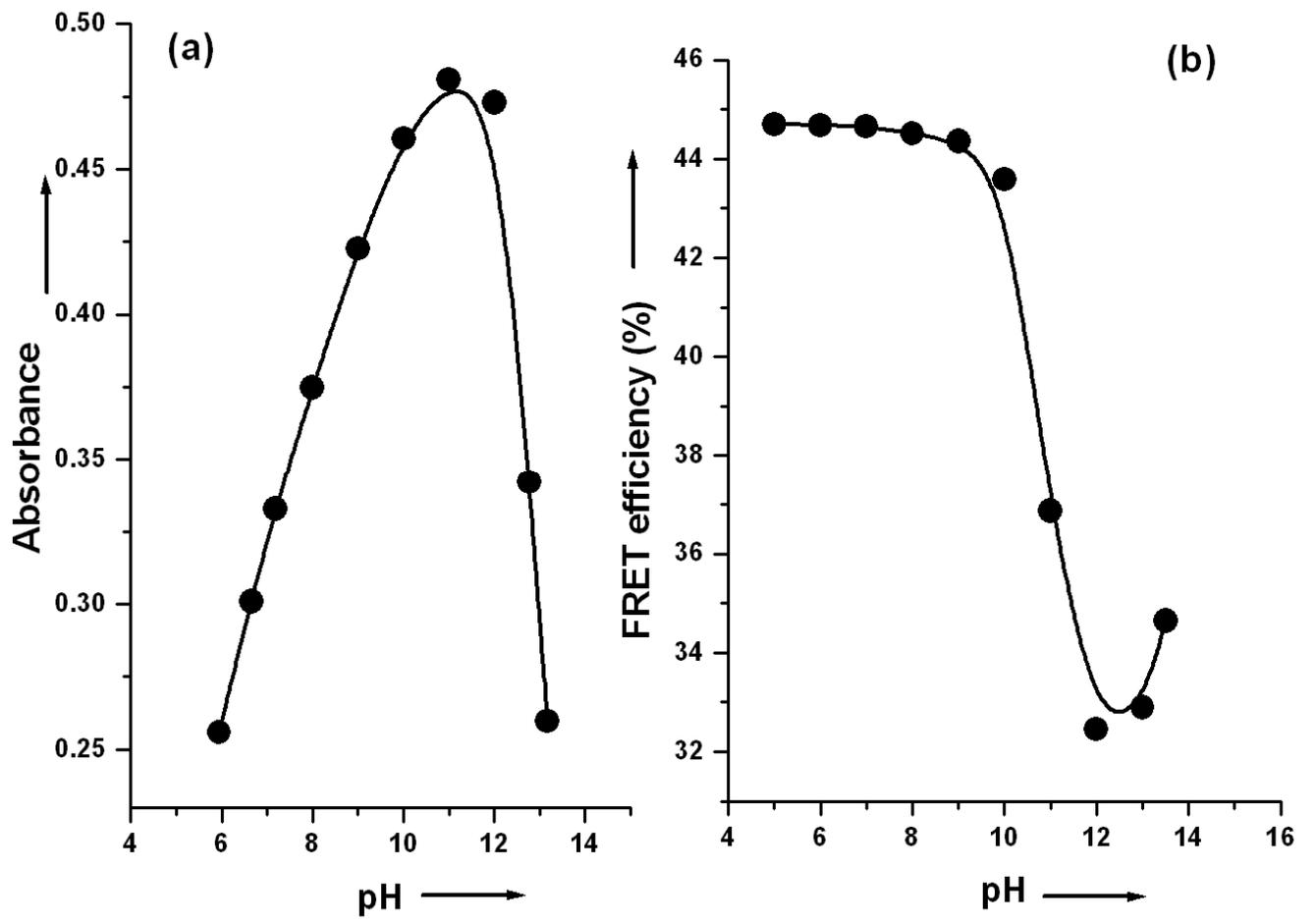

**Fig.2.**

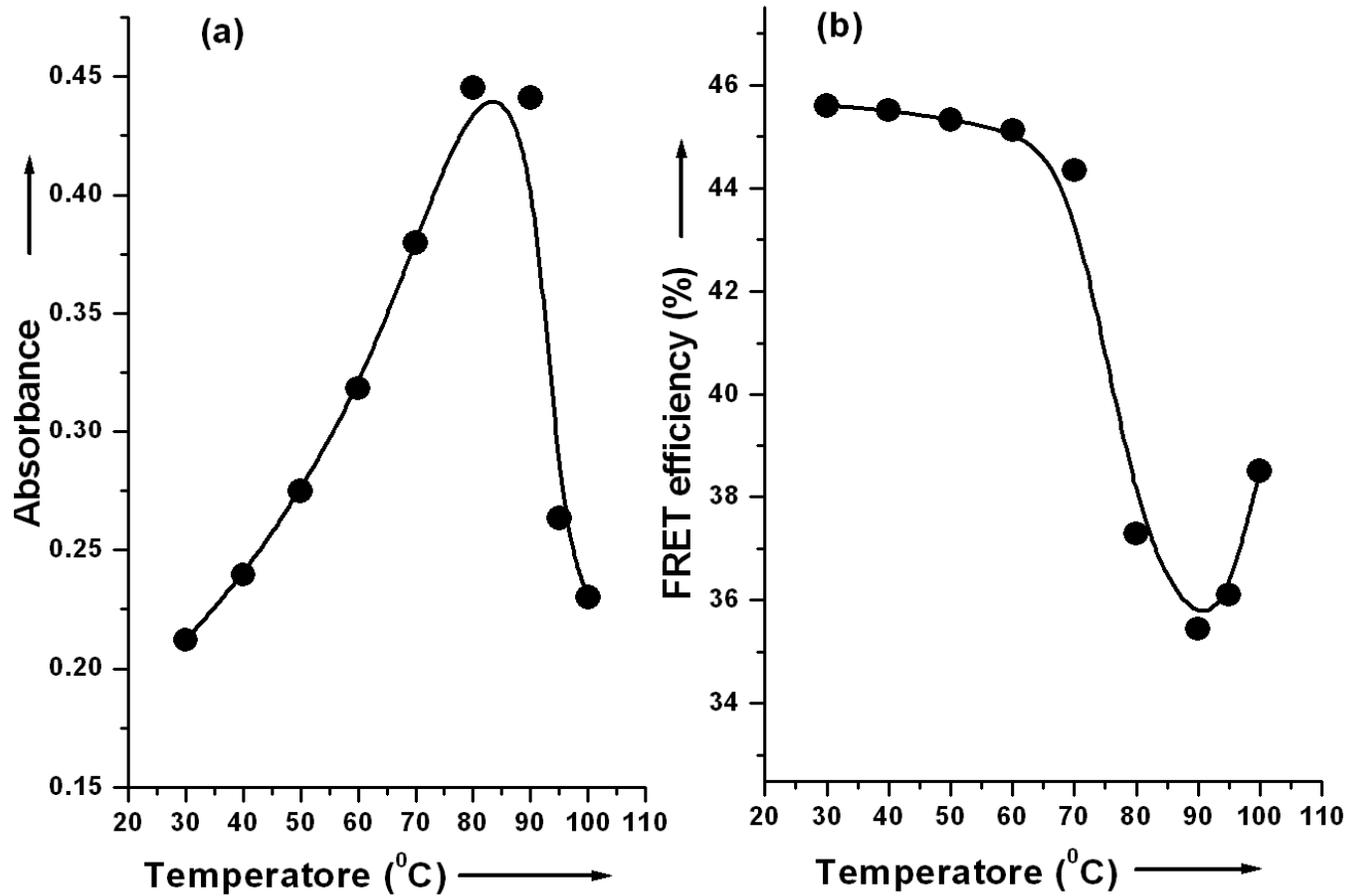

**Fig.3.**

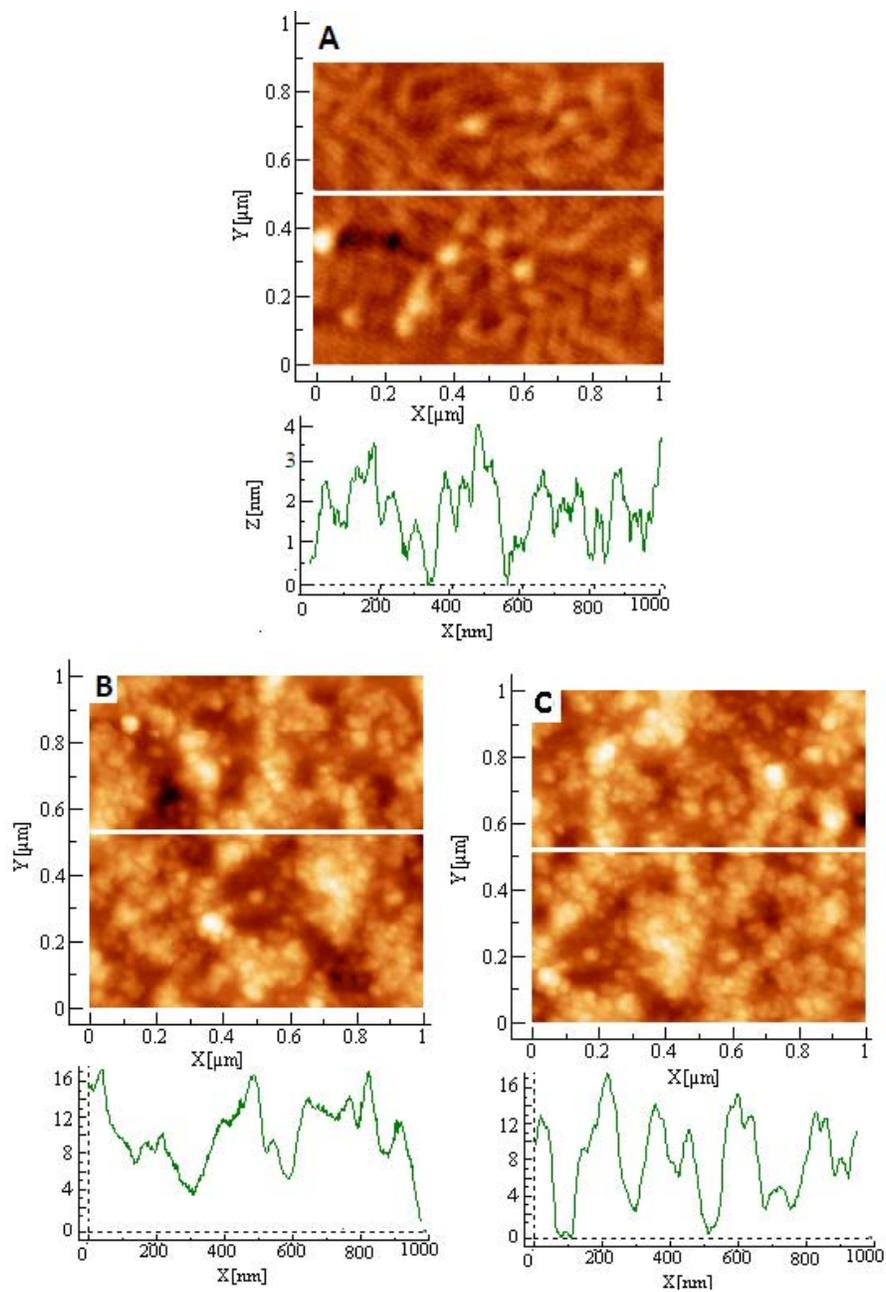

**Fig.4.**

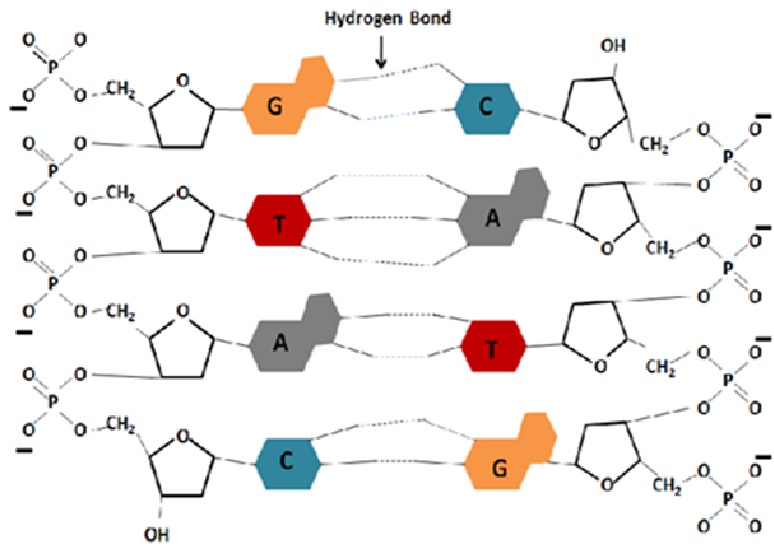
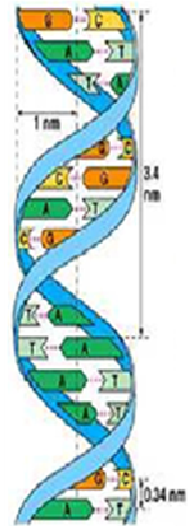

(a) Structure of DNA (showing the four basic components)　　(b) Double helix structure of DNA

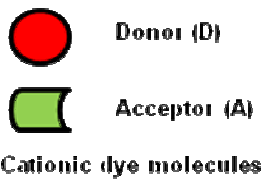
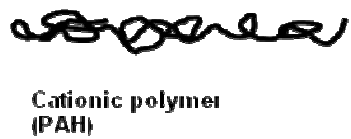
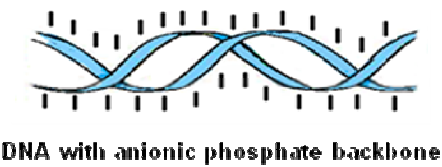

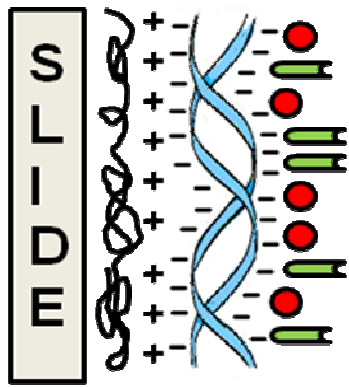
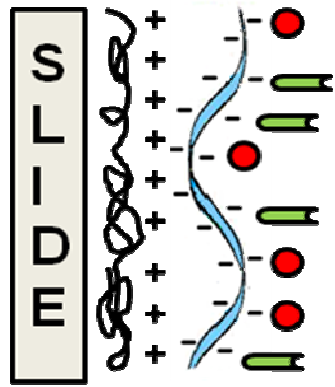
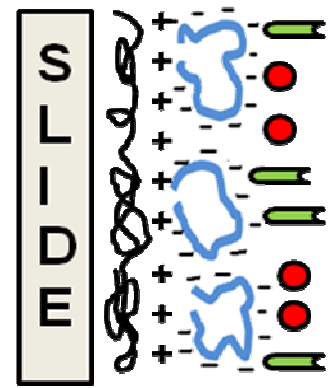

(c) Maximum FRET with double stranded DNA (minimum separation between D and A)

(d) Maximum FRET with denatured DNA (maximum separation between D and A)

(e) FRET increases with coil DNA (separation between D and A decreases)

**Fig.5.**

**Authors Biography:**

**Prof. D. Bhattacharjee** (M.Sc, Kalyani University & Ph.D, IACS, India) is a Professor in the Department of Physics, Tripura University, India. His major fields of interest are preparation and characterization of ultra thin films by Langmuir-Blodgett & Self-assembled techniques. He visited Finland and Belgium for postdoctoral research. He has undertaken several research projects. He has published more than 66 research papers in different national and international journals and attended several scientific conferences in India and abroad.

**Prof. Sangram Sinha** (M.Sc., 1978, Ph.D. 1984) is a Professor in the Department of Botany, Tripura University, India. His major fields of interest are sex determination in dioecious plants; DNA barcoding and molecular karyotyping to assess diversity at inter and intra-genic level / Cytogenetics. He has undertaken several research projects. He has published more than 16 research papers in different national and international journals and attended several scientific conferences in India and abroad.

**Dr. S. A. Hussain** (M.Sc. 2001 & Ph.D, 2007, Tripura University, India) is an Assistant Professor in the Department of Physics, Tripura University. His major fields of interest are Thin Films and Nanoscience. He was a Postdoctoral Fellow of K.U. Leuven, Belgium (2007-08). He received Jagdish Chandra Bose Award 2008-2009, TSCST, Govt. of Tripura; Young Scientist Research Award by DAE, Govt of India. He has undertaken several research projects. He has published 56 research papers in international journals and attended several scientific conferences in India and abroad.

**Dr. P. K. Paul** (M. Sc. Jadavpur University, PhD, Tripura University) is an Assistant Professor in the Department of Physics, Jadavpur University. Presently he is working as Visiting Scientist in Osaka University, Japan. He has published more than 15 research papers in different national and international journals and attended several scientific conferences in India and abroad.

**Mr. Dibyendu Dey** (M.Sc 2009, Tripura University, India) is working as a Research scholar in Department of Physics, Tripura University. His major fields of interest are Fluorescence Resonance Energy Transfer in solution & ultrathin films. He has published 5 research papers in international journals and attended several scientific conferences in India.

**Miss. Jaba Saha** (M.Sc 2012, Tripura University, India) is working as a Research scholar in Department of Physics, Tripura University. Her major fields of interest are Fluorescence Resonance Energy Transfer in solution & ultrathin films and their sensing applications. She has published 1 research papers in international journals and attended several scientific conferences in India.

**Mr Arpan Datta Roy** (M.Sc 2013, Tripura University, India) is working as a Research scholar in Department of Physics, Tripura University. His major fields of interest are study of biomolecules using Fluorescence Resonance Energy Transfer. He has published 1 research papers in international journals and attended several scientific conferences in India.

**Mr. Santanu Chakraborty** (M.Sc 2010, Tripura University, India) is working as a Research scholar in Department of Physics, Tripura University. His major fields of interest are spectroscopic study of organic molecules onto ultrathin films. He has published 4 research papers in international journals and attended several scientific conferences in India and abroad.